# V.V.Sidorenko, S.A.Skorokhod

# The Tippe Top Dynamics: The Comparison of Friction Models

## Equations of motion

Let us consider a spherically shaped top of radius R placed on horizontal plane. The mass distribution inside the top is axially symmetrical. The mass center G of the top lies at a distance a from the geometrical center of the top's surface. Moreover, the contact of sphere and plane has circle's form of radius $\sigma$. At the contact point P the normal reaction force $n$, the force of friction $\vec{f}$ and the torque of friction forces relative to center of contact patch $\vec{m}^r$ are applied to the top.

Following models of friction are considered: viscous, "dry" and empirical friction model (Contesou-Guravlev model [1],[2]) in which pressure distribution in contact patch of body and plane is investigated.

| Friction model | Contact patch's type | Sliding friction force $\vec{f}$ | Friction torque in contact patch $\vec{m}^r$ |
|---|---|---|---|
| Viscous | Point | $-en\vec{\upsilon}_P$ | 0 |
| "Dry" | Point | $-e\dfrac{n\vec{\upsilon}_P}{|\vec{\upsilon}_P|}$ | 0 |
| Contesou-Guravlev | Circle od radius $\sigma$ | $-e\dfrac{n\vec{\upsilon}_P}{|\vec{\upsilon}_P|+8\sigma|\vec{\omega}_\parallel|/3\pi}$ | $-e\dfrac{n\sigma^2\vec{\omega}_\parallel}{5|\vec{\upsilon}_P|+16\sigma|\vec{\omega}_\parallel|/3\pi}$ |

Here $\vec{\upsilon}_P$ is velocity of contact patch's center, $\vec{\omega}_\parallel$ is angular velocity about vertical line, $e$ is constant of friction and $e<<1$ for all models.

To describe the equations of motion we introduce several Cartesian coordinate systems. The system $OXYZ$ is a spatially fixed coordinate system with the axis $OZ$ directed upward; the plane $OXY$ coincides with $\Pi$. The coordinate system $GXYZ$ is originated in the top's center of mass; the axes $GX$, $GY$, $GZ$ are parallel to to the axes $OX$, $OY$, $OZ$ respectively. The coordinate system $G\xi\eta\zeta$ is fixed in a top's body; the axis $G\zeta$ is directed along the symmetry axis. The fixed coordinate system orientation with respect to the system $GXYZ$ is defined by means of Euler's angles $\psi$, $\vartheta$, $\varphi$ ( Fig.2 ). When $\varphi=0$, the fixed coordinate system coincides with a semifixed system $Gxyz$. The axis $Gx$ of the semifixed system is parallel to the plane $\Pi$, the axis $Gz$ coincides with the axis $G\zeta$.

By using the Euler's angles $\psi$, $\vartheta$, $\varphi$ and the coordinates $X_G$, $Y_G$ of the top's center of mass in the coordinate system $OXYZ$, we completely define the position of a top on the plane. It should be note that in case under consideration when the permanent contact of the top with the plane takes place, we have $Z_G = R - a\cos\vartheta$.

Dynamical equations for the top on the plane are a combination of the equations for the motion of the mass center and the equations for the motion of the top about its center of mass. The motion of the mass center is described by the equations

$$m\frac{d\upsilon_{GX}}{dt} = f_X, \qquad m\frac{d\upsilon_{GY}}{dt} = f_Y. \qquad (1.1^1)$$

Here m is the mass of the top, $\upsilon_{GX}$ and $\upsilon_{GY}$ are the components of the mass center velocity in the system $OXYZ$, $f_X$ and $f_Y$ are the components of the sliding force of friction in the same system. In accordance with the accepted assumption about the character of the friction we have following expressions for $f_X$ and $f_Y$:

|  | Viscous friction | "Dry" friction | Contesou-Guravlev friction |
|---|---|---|---|
| $f_X$ | $-en\upsilon_{PY}$ | $-e\dfrac{n\upsilon_{PX}}{|\vec{\upsilon}_P|}$ | $-e\dfrac{n\upsilon_{PX}}{|\vec{\upsilon}_P|+8\sigma|\vec{\omega}_\parallel|/3\pi}$ |
| $f_Y$ | $-en\upsilon_{PX}$ | $-e\dfrac{n\upsilon_{PY}}{|\vec{\upsilon}_P|}$ | $-e\dfrac{n\upsilon_{PY}}{|\vec{\upsilon}_P|+8\sigma|\vec{\omega}_\parallel|/3\pi}$ |

The magnitude of the normal reaction force $n$ can be expressed as:
$$n = m\left[g + a\left(\omega_x^2 \cos\vartheta + \dot{\omega}_x \sin\vartheta\right)\right]$$
$$\upsilon_{PX} = \upsilon_{GX} - \cos\psi\left[(R\cos\vartheta - a)\omega_y - R\omega_z \sin\vartheta\right] - \sin\psi(R - a\cos\vartheta)\omega_x \qquad (1.3)$$
$$\upsilon_{PY} = \upsilon_{GY} - \sin\psi\left[(R\cos\vartheta - a)\omega_y - R\omega_z \sin\vartheta\right] - \cos\psi(R - a\cos\vartheta)\omega_x$$

in which $\omega_x$, $\omega_y$, $\omega_z$ are the projections of the top's angular velocity vector onto the axes of the semifixed coordinate system $Oxyz$ and g is constant of gravity.

The equations for the motion of the top about the center of mass have the following form:

$$\left(A + ma^2 \sin^2\vartheta\right)\frac{d\omega_x}{dt} = -\left(C\omega_z - A\omega_y \operatorname{ctg}\vartheta\right)\omega_y - mga\left[1 + a\omega_x^2/g\right] + \varepsilon m_x$$
$$A\frac{d\omega_y}{dt} = \left(C\omega_z - A\omega_y \operatorname{ctg}\vartheta\right)\omega_x + \varepsilon m_y, \quad C\frac{d\omega_z}{dt} = \varepsilon m_z \qquad (1.1^2)$$
$$\frac{d\psi}{dt} = \omega_y/\sin\vartheta, \quad \frac{d\vartheta}{dt} = \omega_x, \quad \frac{d\varphi}{dt} = \omega_z - \omega_y \operatorname{ctg}\vartheta$$

Here C and A are axial and central transverse moments of inertia, $m_x$, $m_y$, $m_z$ are the projections of the torque caused by the friction force onto the axes of the semifixed system and they depends on the friction model.

In cases of using of dry and viscous friction models the system of equations have Jellet's integral:

$$l = RA\omega_y \sin\vartheta + (R\cos\vartheta - a)C\omega_z = C_0$$

## *Dimensionless procedure*

Let us consider dimensionless variables and parameters. Take as independent variables $\tau$, $\Omega$, $V$, $\rho$ that can be expressed as:

$$\tau = t\sqrt{\frac{mga}{A}}, \quad \Omega = \omega/\sqrt{\frac{mga}{A}}, \quad V = \upsilon\Big/\left(R\sqrt{\frac{mga}{A}}\right), \quad \rho = \sigma/R,$$

Then the parameters $\lambda$, $\alpha$, $\mu$ can be expressed as:

$$\lambda = C/A, \quad \alpha = a/R, \quad \mu = mR^2/A.$$

Taking into account the assumptions made above, we can write the equations of motion in the following form:

$$\mu \dot{V}_{GX} = -\varepsilon F_X,$$
$$\mu \dot{V}_{GY} = -\varepsilon F_Y$$
$$(1+\mu\alpha^2 \sin^2 \vartheta)\dot{\Omega}_x = -(\lambda\Omega_z - \Omega_y \operatorname{ctg} \vartheta)\Omega_y - \sin \vartheta[1+\mu\alpha^2 \Omega_x^2 \cos \vartheta] + \varepsilon M_x \quad (2.1)$$
$$\dot{\Omega}_y = (\lambda\Omega_z - \Omega_y \operatorname{ctg} \vartheta)\Omega_x + \varepsilon M_y, \quad \lambda\dot{\Omega}_z = \varepsilon M_z$$
$$\dot{\psi} = \Omega_y / \sin \vartheta, \quad \dot{\vartheta} = \Omega_x, \quad \dot{\varphi} = \Omega_z - \Omega_y \operatorname{ctg} \vartheta,$$

Here dots mean derivatives with respect to dimensionless time $\tau$, $M_x$, $M_y$, $M_z$ are the dimensionless components of the friction's torque $\vec{M}$ with respect to the center of mass, which may be represented as $\vec{M} = \vec{M}^f + \vec{M}^r$, and $F_X$, $F_Y$ are the dimensionless components of sliding friction force. The corresponding expressions have form:

|  | Viscous friction | "Dry" friction | Contesou-Guravlev friction |
|---|---|---|---|
| $F_X$ | $NV_{PX}\dfrac{1}{\alpha}$ | $N\dfrac{V_{PX}}{|\vec{V}_P|}\dfrac{1}{\alpha}$ | $N\Pi^f V_{PX}\dfrac{1}{\alpha}$ |
| $F_Y$ | $NV_{PY}\dfrac{1}{\alpha}$ | $N\dfrac{V_{PY}}{|\vec{V}_P|}\dfrac{1}{\alpha}$ | $N\Pi^f \dfrac{1}{\alpha}$ |
| $M_x^f$ | $-(1-\alpha\cos\vartheta)NV_{P*}\dfrac{1}{\alpha}$ | $-\dfrac{V_{P*}(1-\alpha\cos\vartheta)}{|\vec{V}_P|}N\dfrac{1}{\alpha}$ | $-\Pi^f V_{P*}(1-\alpha\cos\vartheta)N\dfrac{1}{\alpha}$ |
| $M_y^f$ | $-(\cos\vartheta-\alpha)NV_{Px}\dfrac{1}{\alpha}$ | $-\dfrac{V_{Px}(\cos\vartheta-\alpha)}{|\vec{V}_P|}N\dfrac{1}{\alpha}$ | $-\Pi^f V_{Px}(\cos\vartheta-\alpha)N\dfrac{1}{\alpha}$ |
| $M_z^f$ | $-\sin\vartheta NV_{Px}\dfrac{1}{\alpha}$ | $-\dfrac{V_{Px}\sin\vartheta}{|\vec{V}_P|}N\dfrac{1}{\alpha}$ | $-\Pi^f V_{Px}\sin\vartheta N\dfrac{1}{\alpha}$ |
| $M_x^r$ | 0 | 0 | 0 |
| $M_y^r$ | 0 | 0 | $-\Pi^r \Omega^r N\dfrac{1}{\alpha}\sin\vartheta$ |
| $M_z^r$ | 0 | 0 | $-\Pi^r \Omega^r N\dfrac{1}{\alpha}\cos\vartheta$ |

where $|\vec{V}_P|$, $V_{P*}$, $V_{Px}$, $\Omega^r$, $N$, $\Pi^r$, $\Pi^f$ may be expressed as:

$$|\vec{V}_P| = \sqrt{V_{PX}^2 + V_{PY}^2}$$
$$V_{PX} = V_{GX} - \cos\psi[(\cos\vartheta - \alpha)\Omega_y - \Omega_z \sin\vartheta] - \sin\psi(1-\alpha\cos\vartheta)\Omega_x$$

$$V_{PY} = V_{GY} - \sin\psi\left[(\cos\vartheta - \alpha)\Omega_y - \Omega_z \sin\vartheta\right] - \cos\psi(1 - \alpha\cos\vartheta)\Omega_x$$

$$V_{P*} = -V_{GX}\sin\psi + V_{GY}\cos\psi + (1 - \alpha\cos\vartheta)\Omega_x$$

$$V_{Px} = -V_{GX}\cos\psi + V_{GY}\sin\psi - (\cos\vartheta - \alpha)\Omega_y + \Omega_z \sin\vartheta$$

$$\Omega^r = \Omega_z \cos\vartheta + \Omega_y \sin\vartheta$$

$$N = 1 + \mu\alpha^2\left(\Omega_x^2 \cos\vartheta + \dot{\Omega}_x \sin\vartheta\right)$$

$$\Pi^f = \frac{1}{\left|\vec{V}_P\right| + 8\rho\left|\vec{\Omega}^r\right|/3\pi}$$

$$\Pi^r = \frac{\rho^2}{5|\vec{V}_P| + 16\rho|\vec{\Omega}^r|/3\pi}$$

## *Special evolutionary variables*

First let us consider some properties of the unperturbed motion. For $\varepsilon = 0$ system (2.1) describes the motion of the top along an absolutely smooth surface and has the first integrals

$$V_{GX} = C_1, \ V_{GY} = C_2 \tag{3.1a}$$

$$u \equiv \lambda\Omega_z = C_3, \ \upsilon \equiv \Omega_y \sin\vartheta + \Omega_z \cos\vartheta = C_4 \tag{3.1b}$$

$$E = \frac{1}{2}\left[\mu\left(V_{GX}^2 + V_{GY}^2\right) + (1 + \mu\alpha^2 \sin^2\vartheta)\Omega_x^2 + \Omega_y^2 + \lambda\Omega_z^2\right] - \cos\vartheta = C_5$$

Here $u$ and $\upsilon$ are the projections of the angular momentum onto the axis of symmetry of the top and onto the vertical axis respectively, E denotes the total energy of the top.

In the unperturbed case subsystem (2.1) governing the rotational motion of the top reduces to the form

$$(1 + \mu\alpha^2 \sin^2\vartheta)\dot{\Omega}_x = -(\lambda\Omega_z - \Omega_y \operatorname{ctg}\vartheta)\Omega_y - \sin\vartheta[1 + \mu\alpha^2\Omega_x^2 \cos\vartheta], \tag{3.2}$$

$$\dot{\Omega}_y = (\lambda\Omega_z - \Omega_y \operatorname{ctg}\vartheta)\Omega_x, \quad \lambda\dot{\Omega}_z = 0$$

$$\dot{\psi} = \Omega_y / \sin\vartheta, \qquad \dot{\vartheta} = \Omega_x, \qquad \dot{\varphi} = \Omega_z - \Omega_y \operatorname{ctg}\vartheta,$$

Equations (3.2) are integrable by quadratures [10,11]. In general, in the unperturbed motion, the quantity $\Omega_z$ is constant, $\Omega_x$, $\Omega_y$ and $\vartheta$ are periodic functions $\tau$ with period $T_\vartheta$, $\psi$ and $\varphi$ can be expressed as follows:

$$\psi = \omega_\psi \tau + \psi_1(\tau), \ \varphi = \omega_\varphi \tau + \varphi_1(\tau),$$

Here $\psi_1$ and $\varphi_1$ are $T_\vartheta$-periodic functions of $\tau$. The frequencies $\omega_\vartheta = 2\pi/T_\vartheta$, $\omega_\psi$ and $\omega_\varphi$ depend in a complicated manner on the values of the first integrals (3.1b) and in general are incommensurable.

System (3.2) has a two-parameter family of stationary solutions

$$\Omega_x \equiv 0, \ \Omega_y \equiv \Omega_{y0}, \ \Omega_z \equiv \Omega_{z0}, \ \vartheta \equiv \Theta$$

$$\psi = Wt + \psi_0, \ \varphi = \omega_{\varphi 0}t + \varphi_0. \tag{3.3}$$

The constants $\psi_0$ and $\varphi_0$ in (3.3) are arbitrary, while $\Omega_{y0}$, $\Omega_{z0}$, $\omega_{\varphi 0}$, $W$ and $\Theta$ are connected by the relations:

$$\Omega_{y0} = W \sin\Theta, \qquad \lambda\Omega_{z0} = -\frac{1}{W} + W\cos\Theta,$$

$$\omega_{\varphi 0} = \frac{1}{\lambda C}\left[(\lambda - 1)W\cos\Theta - \frac{1}{W}\right]$$

Solutions (3.3) correspond to those motions which can be represented by a certain superposition of a uniform rotation about the axis of symmetry and a uniform rotation about the vertical. Such motions are called "regular precessions". It is convenient to choose the velocity of the precession $W$ and the angle of nutation $\Theta$ as the parameters of the family (3.3).

A closed subsystem of equations for $\Omega_x$, $\Omega_y$ and $\vartheta$ can be derived from (3.3), containing $\Omega_z$ as a parameter. Setting

$$\Omega_z = \frac{1}{\lambda}\left(-\frac{1}{W} + W\cos\Theta\right) \tag{3.4}$$

we consider an integral manifold $S_{W,\Theta}$ in the phase space $(\Omega_x, \Omega_y, \vartheta)$ with a fixed value for the integral $\upsilon$, pertaining to the regular precession with the parameters $W$ and $\Theta$ [12]. Its parametric representation has the form

$$S_{W,\Theta} = \{(\Omega_x, \Omega_y, \vartheta): \Omega_x = \Omega_x(W, \Theta, c, v), \Omega_y = \Omega_y(W, \Theta, c, v), \vartheta = \vartheta(W, \Theta, c, v);$$
$$0 \leq v \leq 2\pi, 0 \leq c \leq c_0(W, \Theta)\}$$

where $c$ and $v$ denote the amplitude and the phase of the nutational oscillations. At individual solution lying on the manifold $S_{W,\Theta}$ $v = \omega_\Theta \tau + v_0$. It is not difficult to prove, through Lyapunov's holomorfic integral theorem [13], that the functions $\Omega_x = \Omega_x(W,\Theta,c,v)$, $\Omega_y = \Omega_y(W,\Theta,c,v)$, $\vartheta = \vartheta(W,\Theta,c,v)$ can be written in the form of the series

$$\Omega_x = \sum_{k=1}^{\infty} c^k \Omega_{xk}(W,\Theta,v), \qquad \Omega_y = \Omega_{y0}(W,\Theta) + \sum_{k=1}^{\infty} c^k \Omega_{yk}(W,\Theta,v),$$

$$\vartheta = \Theta + \sum_{k=1}^{\infty} c^k \vartheta_k(W,\Theta,v) \tag{3.5}$$

which converge for sufficiently small values of $|c|$ ( to apply Lyapunov's theorem it is necessary to reduce the order of the system for $\Omega_x$, $\Omega_y$ and $\vartheta$ using the integral $\upsilon$).

We have the following expressions for the first coefficients

$$\Omega_{x1} = -\omega_0 \sin v, \quad \Omega_{y1} = -\cos v / W, \quad \vartheta_1 = \cos v$$

Here $\omega_0 = \sqrt{\dfrac{W^4 + 2W^2 \cos\Theta + 1}{W^2(1 + \mu\alpha^2 \sin^2\Theta)}}$ - is the frequency of the small nutational oscillations.

The formulae (3.4), (3.5) define the local change of variables

$$(\Omega_x, \Omega_y, \Omega_z, \vartheta) \to (W, \Theta, c, v)$$

The new variables have a simple mechanical meaning: $W$ and $\Theta$ specify the reference regular precession, while $c$ and $v$ characterize the amplitude and phase of the nutational oscillations in motion which is close to the reference precession. It is implied that this motion and the reference precession belong to the same joint level of the integrals $u$ and $\upsilon$.

This change of variables reduces system (2.1) to a form which is convenient for the application of the averaging method [14].

Variables $W$ $\Theta$ and $c$ are independent integrals of unperturbed system. The following relations hold

$$u = -\frac{1}{W} + W\cos\Theta, \quad \upsilon = -\frac{\cos\Theta}{W} + W \tag{3.6}$$

## Equations of motion of the top in the special evolutionary variables

At first, we obtain equations for the variables $W$, $\Theta$ by means of two sequential substitutions:
$$(\Omega_y, \Omega_z) \to (u, \upsilon) \to (W, \Theta).$$

For $\varepsilon \neq 0$ the change in the projections of the angular momentum onto the symmetry axis and onto the vertical is described by the equations

$$\dot{u} = M_z, \quad \dot{\upsilon} = M_z \cos\vartheta + M_y \sin\vartheta \qquad (4.1)$$

Expressing $u$, $\upsilon$ in (4.1) in terms of $W$ and $\Theta$ in accordance with (3.6), we find

$$\frac{\partial u}{\partial W}\dot{W} + \frac{\partial u}{\partial \Theta}\dot{\Theta} = \varepsilon\left(M_z^f + M^r \cos\vartheta\right)$$
$$\frac{\partial \upsilon}{\partial W}\dot{W} + \frac{\partial \upsilon}{\partial \Theta}\dot{\Theta} = \varepsilon\left(M_z^f \cos\vartheta + M_y^f \sin\vartheta + M^r\right) \qquad (4.2)$$

Equations (4.2) define a system of linear equations for $\dot{W}$ and $\dot{\Theta}$ with the determinant

$$D = \frac{\partial(u, \upsilon)}{\partial(W, \Theta)} = \omega_0^2 \sin\Theta(1 + \mu\alpha^2 \sin\Theta)/W$$

System (4.2) can be solved if $W \neq 0$ and $\sin\Theta \neq 0$:

$$\dot{W} = -\varepsilon\left[\sin\Theta F_x \frac{\partial L}{\partial \Theta} - M^r\left(\cos\vartheta \frac{\partial \upsilon}{\partial \Theta} - \frac{\partial u}{\partial \Theta}\right)\right] \cdot \frac{\partial(W, \Theta)}{\partial(u, \upsilon)}$$
$$\dot{\Theta} = \varepsilon\left[\sin\Theta F_x \frac{\partial L}{\partial W} - M^r\left(\cos\vartheta \frac{\partial \upsilon}{\partial W} - \frac{\partial u}{\partial W}\right)\right] \cdot \frac{\partial(W, \Theta)}{\partial(u, \upsilon)} \qquad (4.3)$$

where $L = \upsilon - \alpha u$, and the expressions for $F_x$ и $M^r$ are listed below:

|  | Viscous friction | "Dry" friction | Contesou-Guravlev friction |
|---|---|---|---|
| $F_x$ | $NV_{Px}\dfrac{1}{\alpha}$ | $\dfrac{V_{Px}}{|\vec{V}_P|} N \dfrac{1}{\alpha}$ | $\Pi^f V_{Px} N \dfrac{1}{\alpha}$ |
| $M^r$ | 0 | 0 | $-\Pi^r \Omega^r N \dfrac{1}{\alpha}$ |

Substitution $(\Omega_x, \vartheta) \to (c, \nu)$ is analogous to the Van der Pol substitution [14]. Slightly modifying the Van der Pol approach, we find expression for $c$ (there is no reason for consideration of $\nu$)

$$\dot{c} = -\frac{1}{\Delta_0}\left(\Delta_c^W \dot{W} + \Delta_c^\Theta \dot{\Theta}\right) + \frac{\varepsilon}{\Delta_0} \frac{M_x}{1 + \mu\alpha^2 \sin^2\Theta} \frac{\partial Q}{\partial \nu} \qquad (4.4)$$

Here $Q(W, \Theta, c, \nu) = \vartheta - \Theta$ and functions $\Delta_0$, $\Delta_W^c$, $\Delta_\Theta^c$, $\Delta_W^\nu$, $\Delta_\Theta^\nu$ are defined by formulae

$$\Delta_0 = \frac{\partial \omega_\vartheta}{\partial c}\left(\frac{\partial Q}{\partial \nu}\right)^2 - \omega_\vartheta\left[\frac{\partial Q}{\partial c}\frac{\partial^2 Q}{\partial \nu^2} - \frac{\partial Q}{\partial \nu}\frac{\partial^2 Q}{\partial c \partial \nu}\right],$$

$$\Delta_W^c = \frac{\partial \omega_\vartheta}{\partial W}\left(\frac{\partial Q}{\partial \nu}\right)^2 - \omega_\vartheta\left[\frac{\partial Q}{\partial W}\frac{\partial^2 Q}{\partial \nu^2} - \frac{\partial Q}{\partial \nu}\frac{\partial^2 Q}{\partial W \partial \nu}\right] \qquad (4.5)$$

$$\Delta_\Theta^c = \frac{\partial \omega_\vartheta}{\partial \Theta}\left(\frac{\partial Q}{\partial \nu}\right)^2 - \omega_\vartheta\left[\left(1 + \frac{\partial Q}{\partial \Theta}\right)\frac{\partial^2 Q}{\partial \nu^2} - \frac{\partial Q}{\partial \nu}\frac{\partial^2 Q}{\partial \Theta \partial \nu}\right]$$

## *Averaged equations in the case of viscous friction*

In the first approximation of the averaging method we find:

$$\dot{V}_{GX} = -\varepsilon \frac{V_{GX}}{\mu\alpha} + O(\varepsilon^2), \quad \dot{V}_{GY} = -\varepsilon \frac{V_{GY}}{\mu\alpha} + O(\varepsilon^2)$$

$$\dot{W} = -\varepsilon \sin\Theta U \frac{dL}{d\Theta} \frac{1}{\alpha} \frac{\partial(W,\Theta)}{\partial(u,v)} + O(\varepsilon^2) \quad (5.1)$$

$$\dot{\Theta} = \varepsilon \sin\Theta U \frac{dL}{dW} \frac{1}{\alpha} \frac{\partial(W,\Theta)}{\partial(u,v)} + O(\varepsilon^2)$$

$$\dot{c} = \varepsilon c \left[ \Xi_1 + (\Xi_2 + U\Xi_3) \frac{\partial(W,\Theta)}{\partial(u,v)} \right] + O(\varepsilon^2)$$

Here $U = \sin\Theta[\Omega_{z0}(W,\Theta) - (\cos\Theta - \alpha)W]$ - is the averaged projection of absolute velocity of the point P on the Gx axis in the regime of regular precession of the top at a velocity $W$ with a nutation angle $\Theta$, functions $\Xi_1(\Theta), \Xi_2(W,\Theta), \Xi_3(W,\Theta)$ are given by

$$\Xi_1 = -\frac{(1-\alpha\cos\Theta)^2}{2\alpha(1+\mu\alpha^2\sin^2\Theta)}$$

$$\Xi_2 = \frac{\sin\Theta}{\alpha}\left[\frac{L}{2} - (1-\alpha\cos\Theta)W\right]\frac{\partial L}{\partial W}$$

$$\Xi_3 = \frac{\sin\Theta}{4\alpha\omega_0^2}\frac{\partial(\omega_0^2,L)}{\partial(W,\Theta)} - \left[\cos\Theta - \frac{\mu\alpha^2}{2}\omega_0^2\sin^2\Theta\right]\frac{1}{\alpha}\frac{\partial L}{\partial W}$$

As well as the original system (2.1), the averaged equations in the case of viscous friction have Jellet's integral:

$$L = v - \alpha u = C_0.$$

The fact of lack of interaction between top's center of mass motion and angular motion allows us to express phase portraits on the plain $(W,\Theta)$ with regions of increase/decrease of magnitude of $c$. As an example, the figures drawn below show phase portraits for a top with parameters $r/a = 5$, $mr^2/A = 2.25$ and C/A = 0.6, 0.9, 1.1, 1.5 respectively, the regions of increase of $c$ are shaded.

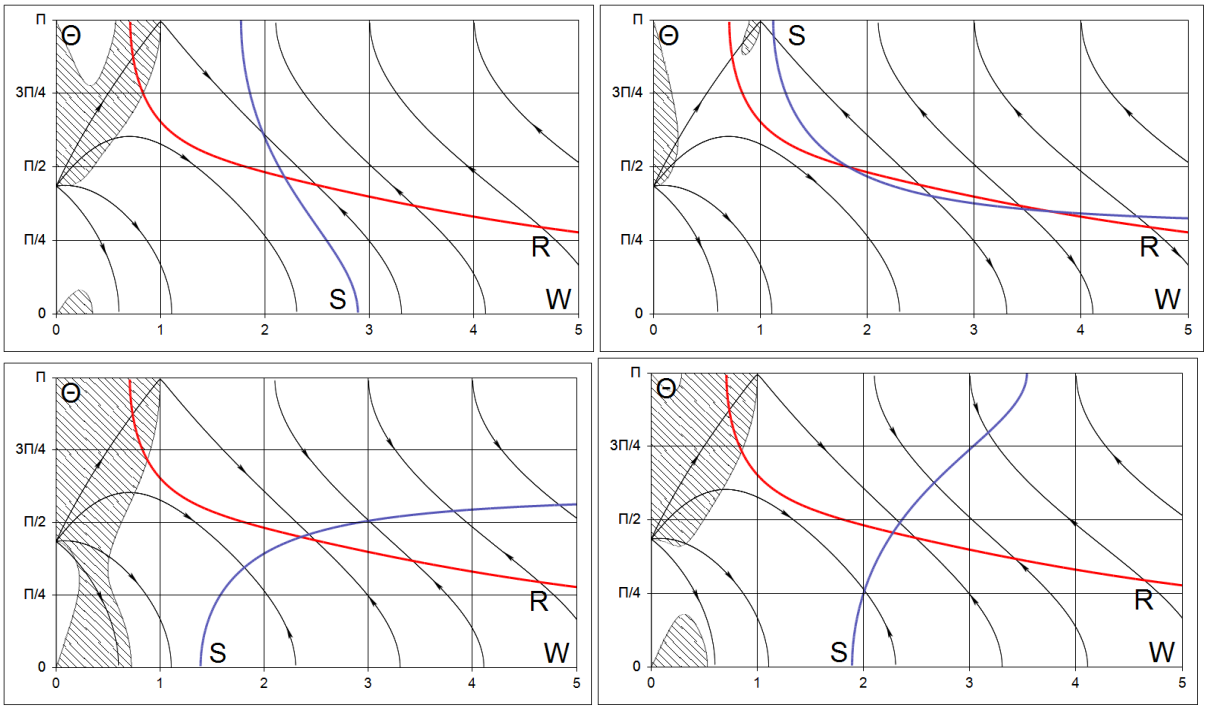

## *Averaged equations in the case of "dry" friction*

In the first approximation of the averaging method applied on the integral manifold $V_{GX} = V_{GY} = 0$ we find:

$$\dot{W} = -\varepsilon \sin\Theta \frac{U}{|U|} \frac{dL}{d\Theta} \frac{1}{\alpha} \frac{\partial(W,\Theta)}{\partial(u,\upsilon)} + O(\varepsilon^2)$$

$$\dot{\Theta} = \varepsilon \sin\Theta \frac{U}{|U|} \frac{dL}{dW} \frac{1}{\alpha} \frac{\partial(W,\Theta)}{\partial(u,\upsilon)} + O(\varepsilon^2) \qquad (6.1)$$

$$\dot{c} = \frac{\varepsilon c}{|U|} \left[ \Xi_1 + U\Xi_3^* \frac{\partial(W,\Theta)}{\partial(u,\upsilon)} \right] + O(\varepsilon c^2)$$

Here the meaning of U is the same as in the case of viscous friction, functions $\Xi_1(\Theta)$, $\Xi_3(W,\Theta)$ are given by

$$\Xi_1 = -\frac{(1-\alpha\cos\Theta)^2}{2\alpha(1+\mu\alpha^2 \sin^2\Theta)}$$

$$\Xi_3^* = \frac{\sin\Theta}{4\alpha\omega_0^2} \frac{\partial(\omega_0^2, L)}{\partial(W,\Theta)} - \frac{1}{2\alpha}\left[\cos\Theta - \mu\alpha^2\omega_0^2 \sin^2\Theta\right]\frac{\partial L}{\partial W}$$

As well as the original system (2.1), the averaged equations in the case of viscous friction have Jellet's integral:

$$L = \upsilon - \alpha u = C_0.$$

On phase portraits the regions of increase of $c$ are shaded. As an example, the figures drawn below show phase portraits for a top with parameters $r/a = 5$, $mr^2/A = 2.25$ and C/A = 0.6, 0.9, 1.1, 1.5 respectively. In comparison with the case of viscous friction the regions of increase/decrease of magnitude of $c$ are slightly different.

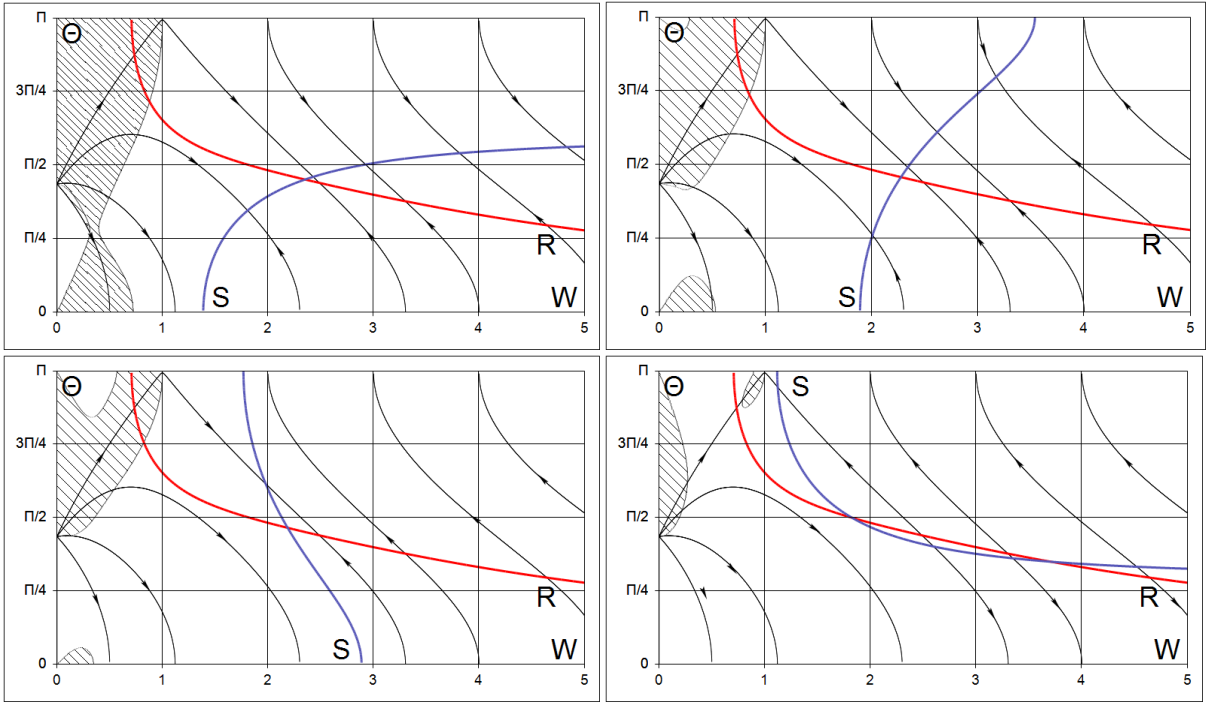

## *Averaged equations in the case of "Contesou-Guravlev" friction*

In the first approximation of the averaging method applied on the integral manifold $V_{GX} = V_{GY} = 0$ we find:

$$\dot{W} = -\varepsilon\left[\sin\Theta \Pi_0^f U \frac{dL}{d\Theta} - \Pi_0^r \Omega_0^r\left(\cos\Theta \frac{\partial \upsilon}{\partial \Theta} - \frac{\partial u}{\partial \Theta}\right)\right]\frac{1}{\alpha}\frac{\partial(W,\Theta)}{\partial(u,\upsilon)} + O(\varepsilon^2)$$

$$\dot{\Theta} = \varepsilon\left[\sin\Theta \Pi_0^f U \frac{dL}{dW} - \Pi_0^r \Omega_0^r\left(\cos\Theta \frac{\partial \upsilon}{\partial W} - \frac{\partial u}{\partial W}\right)\right]\frac{1}{\alpha}\frac{\partial(W,\Theta)}{\partial(u,\upsilon)} + O(\varepsilon^2)$$

$$\dot{c} = \varepsilon c\left\{\Pi_o^f\left[\Xi_1 + (\Xi_2 + U\Xi_3)\frac{\partial(W,\Theta)}{\partial(u,\upsilon)}\right] + \Pi_o^r[\Xi_4 + \Omega_0^r\Xi_5]\frac{\partial(W,\Theta)}{\partial(u,\upsilon)}\right\} + O(\varepsilon^2) \qquad (7.1)$$

Here $\Omega_0^r = W\sin^2\Theta + \Omega_{z0}\cos\Theta$ - is the averaged projection of angular velocity of the top on the vertical direction in the regime of regular precession at a velocity $W$ with a nutation angle $\Theta$, functions $U$, $\Xi_1$, $\Xi_2$, $\Xi_3$ the same as in previous cases, $\Pi_0^f$ and $\Pi_0^r$ are given by

$$\Pi_0^f = \frac{1}{|U| + 8\rho|\Omega^r|/3\pi}$$

$$\Pi_0^r = \frac{\rho^2}{5|U| + 16\rho|\Omega^r|/3\pi},$$

The functions $\Xi_4(W,\Theta)$, $\Xi_5(W,\Theta)$ are not listed on account of they unhandiness.

In contrast to the cases of dry and viscous friction models averaged equations in the case of "Contesou-Guravlev" friction do not have Jellet's integral and his evaluation is given by

$$\dot{L} = -\varepsilon(1 - \alpha\cos\Theta)\frac{1}{\alpha}\Pi_0^r\Omega^r$$

As in the case of "dry" and viscous friction let us consider the phase portraits in the plane $(W,\Theta)$. As an example, the figures drawn below show phase portraits for a top with parameters $r/a = 5$, $mr^2/A = 2.25$, $\sigma/r = 0.07$, $\varepsilon = 0.05$ and C/A = 0.6, 0.9, 1.1, 1.5 respectively

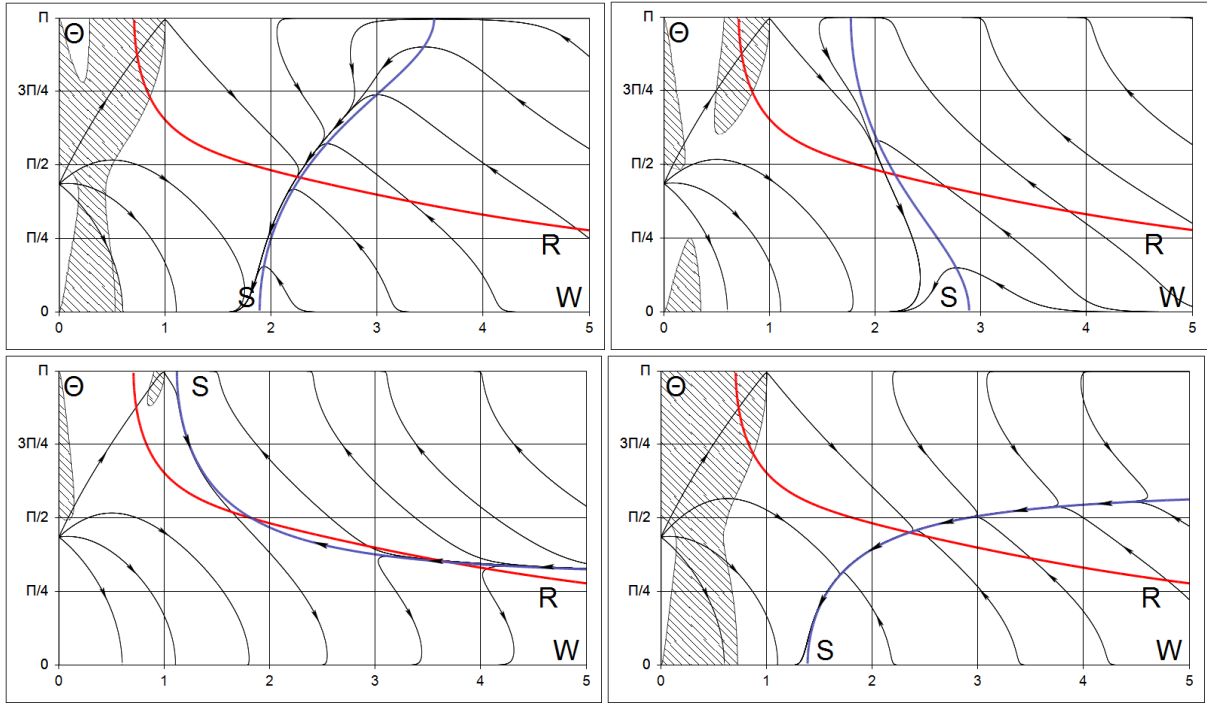

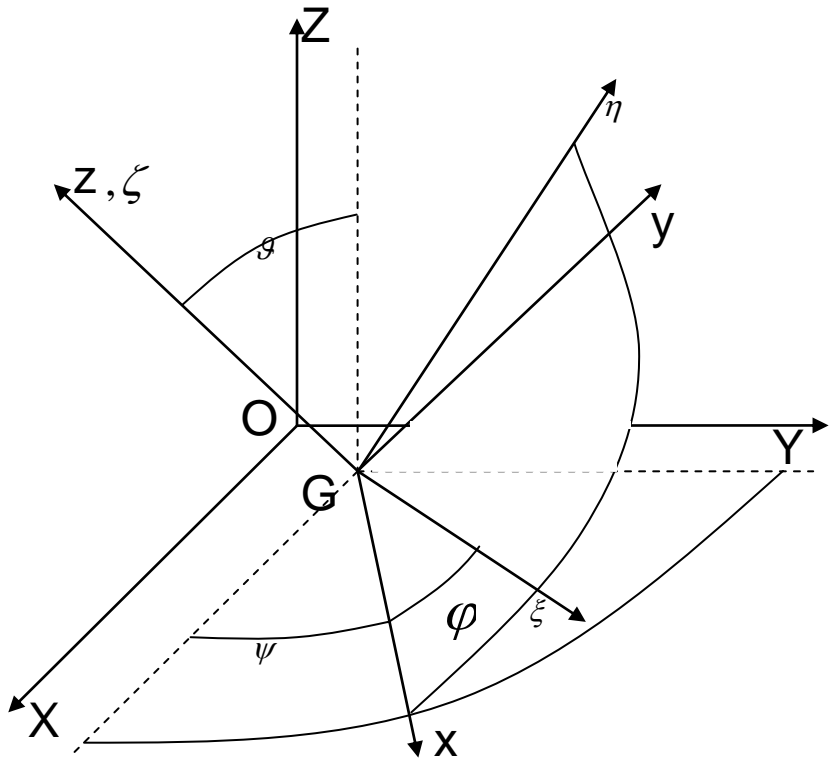

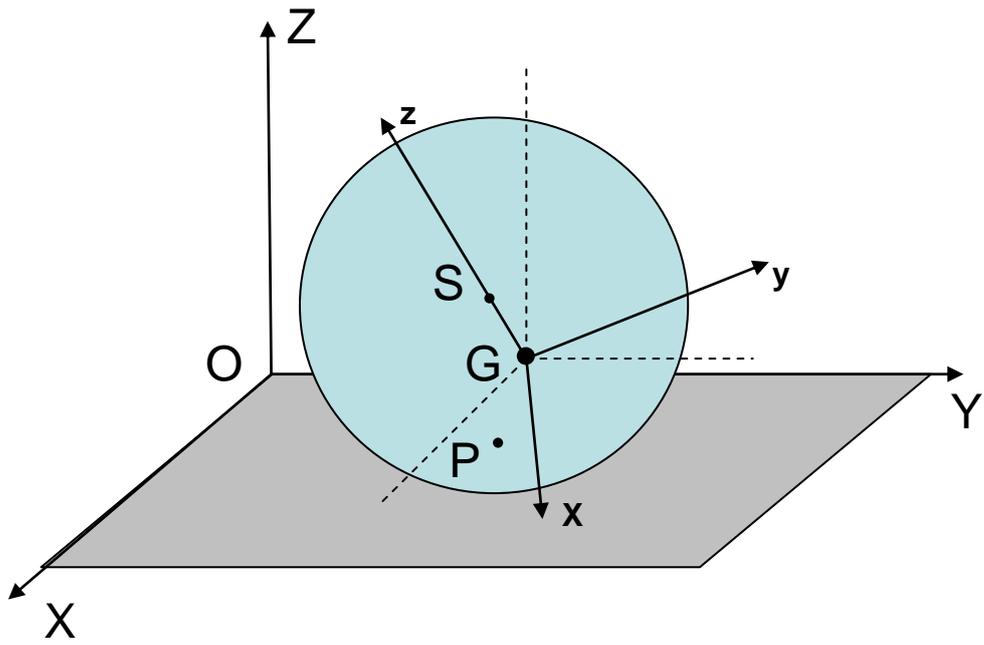